# Laser Communication with Proxima and Alpha Centauri using the Solar Gravitational Lens

Geoffrey W. Marcy[1*], Nathaniel K. Tellis[2], Edward H. Wishnow[3]

[1] *Center for Space Laser Awareness, 3388 Petaluma Hill Rd, Santa Rosa, CA, 95404, USA*
[2] *RocketCDL*
[3] Space Science Laboratory, University of California, Berkeley

ABSTRACT

A search was conducted for laser signals, both sub-second pulses and continuous emission, from the regions of the sky opposite Proxima and Alpha Centauri. These regions are located at the foci of the gravitational lensing caused by the Sun, ideal for amplifying transmissions between our Solar System and those two nearest stellar neighbors. The search was conducted using two objective prism telescopes operating with exposure times of 0.25 seconds, enabling detection of sub-second laser pulses coming from the Solar gravitational foci. During six months in 2020 and 2021, 88000 exposures for Proxima Cen and 47000 exposures for Alpha Cen were obtained.  No evidence was detected of light pulses or continuous laser emission in the wavelength range of 380 to 950 nm. We would have detected a laser having a power of just 100 Watts, for a benchmark 1-meter laser launcher that was diffraction-limited and located at the Sun's gravitational focus 550 AU away. To be detected, that beam must intercept Earth either by intention, by accident, or if intended for a probe near Earth that is communicating with another one at the Solar gravitational lens. These non-detections augment a previous non-detection of laser light coming directly from Proxima Centauri conducted with the HARPS spectrometer on the ESO 3.6-meter telescope.

## 1  INTRODUCTION

The Milky Way Galaxy may contain a network of mutually communicating spacecraft stationed near stars (e.g., Bracewell 1973; Freitas 1980; Maccone 2014, 2021; Gillon 2014, Hippke 2020, 2021ab; Gertz 2018, 2021).  The spacecraft may perform scientific exploration and surveillance using an architecture of communication nodes, enabling collection and dissemination of local observations of the Galaxy, including evolving planetary systems and any biology.  The probes may communicate with privacy, high bandwidth, and minimal onboard payload by using ultraviolet, optical, or infrared laser communication (Schwartz and Townes 1961, Zuckerman 1985, Hippke 2018, 2021ab). Such a laser communication network is already proliferating among satellites orbiting Earth, with a future of quantum efficiency and encryption (Hippke 2021c).

Laser light is nearly monochromatic (e.g., Naderi et al. 2016, Su et al. 2014, Wang et al. 2020), prompting searches for narrowband, optical light using high-resolution spectra of more than 5000 normal stars of spectral type F, G, K, and M, yielding no detections and no viable candidates (Reines & Marcy 2002; Tellis & Marcy 2017).  A similar search for laser emission from more massive stars of spectral types O, B, and A, has also revealed no viable candidates (Tellis et al. 2021, in preparation).  Recently, we performed a search of 107 high-resolution optical spectra of Proxima Centauri for laser emission but found none down to kilowatt power thresholds (Marcy 2021).  These laser searches involved examining high-resolution spectra, $\lambda/\Delta\lambda > 60000$, in the wavelength range $\lambda$ = 3600 to 9500 Å, for nearly monochromatic emission lines. The required laser power for detection is 50 kW to 10 MW, assuming a diffraction-limited laser emitter consisting of a benchmark 10-meter aperture. None was found.

Some searches for laser pulses have been done using broadband light.  Searches for both continuously transmitting and short duration light pulses in the visible spectrum (optical SETI) have been conducted (e.g., Wright et al. 2001; Howard et al. 2004; Stone et al. 2005; Howard et al. 2007, Hanna et al. 2009, Abeysekara et al. 2016, Villarroel et al. 2020, 2021). No optical laser emission has ever been found in the Milky Way Galaxy, nor elsewhere in the universe.  Next generation searches for optical pulses are in progress (Maire et al. 2020).

*E-mail: geoff@spacelaserawareness.org

The power requirement for interstellar communication can be reduced by many orders of magnitude by focusing the transmissions with gravitational lenses of stars, as suggested by von Eshleman 1979, Maccone 2011, 2014, 2021, Gillon 2014, Hippke 2020, 2021ab, and Gertz 2021. The optical properties of a star, including its non-spherical gravitational potential, as a gravitational lens have been calculated precisely (Turyshev & Toth 2021). The transceivers must be positioned anywhere along the gravitational focal line located at least 550 au away (for Solar mass) on the far side of one or both stars (Hippke 2021ab). The focal length of the gravitational lens depends on the closest approach distance of the beam as it passes the star and inversely on the mass of the star. Laser communication from one probe can be directed toward its host star (e.g. the Sun) which gravitationally focuses the beam at the probe located at a nearby star, amplifying the otherwise weak signal intensity by over a million times. In effect, the gravitational focusing of laser communication by stars may serve as the fiber-optic conduit of a Galactic internet.

This million-fold amplification of communication signals motivates a search for spacecraft at the gravitational focus regions of one star by its nearby stars (e.g., Gillon 2014, Hippke 2021, and Gertz 2021, Maccone 2021). The light we detect from such probes there could be either the communication beam itself (if wide enough) or a laser beam for some other purpose including pinging the local habitable planet, e.g. the Earth. Hence, we should search these gravitational lens foci at all wavelengths of the electromagnetic spectrum for signs of such spacecraft.

Figure 1 shows a sketch of the scene. A transmitter on a probe located at the Solar gravitational lens focus at far left, 550 au away, can use the Sun as a gravitational lens to amplify communication with a receiver near Alpha Centauri at far right. The Earth may reside within that communication beam, or it may receive laser light (shown in green) from the probe that is emitted by the probe, intentionally or not. A device may be communicating with a probe located very close to the Earth or indeed attempting to communicate with us. This program is to search for laser light from the probe. The search for extraterrestrial signals has a higher probability of success near the Solar gravitational lens focal region of the sky than from other randomly chosen regions. Figure 1 is not to scale.

In August 2020, we initiated a search for monochromatic optical light from the Solar gravitational lens (SGL) focal regions located in the sky opposite four nearby stars, Tau Ceti, Proxima Centauri, and Alpha Centauri AB. Those stars do not reside within a few degrees of the ecliptic plane, so the Earth is never positioned in the SGL communication beam. However, we search there for inadvertent or intentional optical laser light that may be sent toward Earth. In addition, this search would capture, for free, any laser pulses coming from behind the SGL focal region in a 2-degree cone extending into the Milky Way. The motivation to observe the SGL focal region of the Proxima Centauri system was boosted by an apparent radio signal at 982 MHz discovered while the Murriyang radio telescope was pointed at Proxima Centauri (Sheikh et al. 2021, Smith et al. 2021), though it may have been caused by radio frequency interference from electronics on Earth. This paper describes a search for laser emission, pulsed and continuous, from the SGL focus regions opposite the Sun from Proxima and Alpha Centauri.

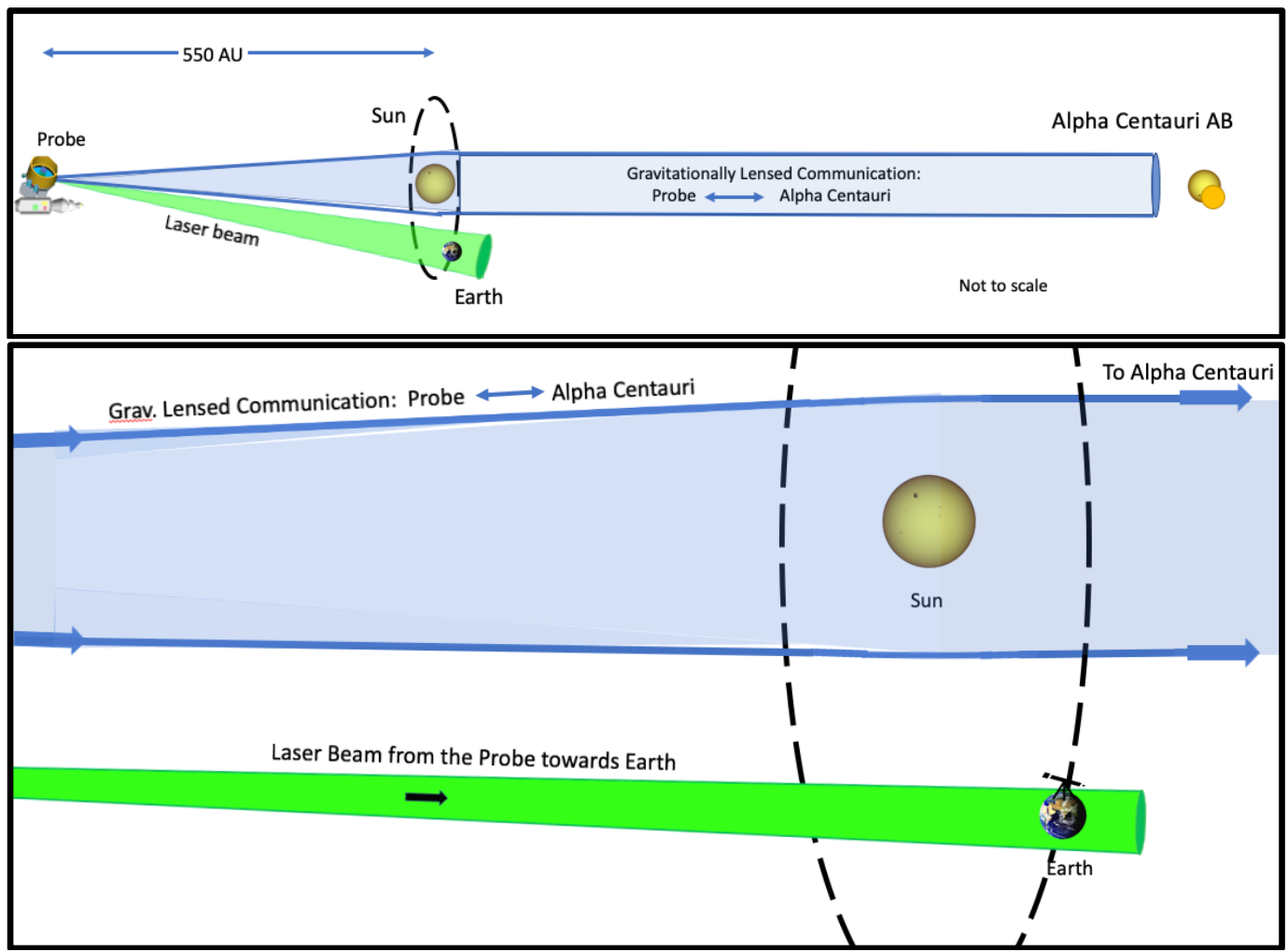

*Figure 1. A sketch of the geometry. Top: The probe is at left, the Sun and Earth in the middle, and Alpha Centauri AB at right (not to scale). Advanced civilizations might place a probe 550 AU to the left of the Sun to utilize its gravitational field as a lens to amplify communication with Alpha Centauri (at right). Communication can go in both directions. Telescopes on Earth may eavesdrop on that communication beam (blue), or may receive a separate laser beam (green) from the probe. Bottom: A close-up of the beams near the Sun and Earth. As the beam widths are unknown, the blue and green beams may overlap if extraterrestrial probes emit suboptimal beam widths for technical trade-offs we can't anticipate.*

## 2  THE DOUBLE OBJECTIVE PRISM TELESCOPE

We designed and constructed an objective prism Schmidt telescope capable of performing low resolution spectroscopy over an entire 2-degree field of view with a time resolution of 0.25 sec, making the system sensitive to monochromatic pulses of optical light. We chose an objective prism to obtain simultaneous spectra over a wide field and to avoid the confusion of additional spectral orders that arise from diffraction gratings, including zeroth-order points of light that mimic monochromatic sources, including glints (Corbett et al. 2021, Nir et al. 2021). While our design was dictated by the desire to search a wide field for monochromatic pulses, the design and spectral resolution are remarkably similar to the historically important objective prism telescopes of the Harvard Observatory (Pickering 1912; Fleming 1917).

The primary telescope is based on the 'Celestron RASA-11' modified Schmidt telescope with an aperture of 0.28 m (11 inch) and a focal ratio of f/2.2. The prism was designed to resolve the [OIII] lines at 5007 and 4959 Å to identify (and reject) regions of ionized gas such as planetary nebulae, HII regions, and active galactic nuclei. The resulting prism has a 7-degree wedge angle with an outer diameter of 0.3 m and a central hole 0.14 m in diameter for the camera. The prism glass type is 'CDGM F4', equivalent to Schott glass F2. The prism has ¼ λ flatness and was fabricated by *Sinoptix*. The prism was mounted at the front of the RASA telescope, upstream of the corrector plate, but supported within the existing light baffle, as shown in Figure 2. The stellar spectra have a dispersion of ~450 Å/mm near 4300 Å, which is, not coincidentally, similar to the low resolution spectra of the Draper and Bache telescopes (Flemming 1917).

We mounted a Fingerlakes KL-400 CMOS camera at prime focus, just outside the prism and correction optics (see Figure 2). The camera has 2048x2048 pixels of 11μm size. The detector quantum efficiency is over 90% from 480 to 650 nm, falling to 10% at wavelengths 380 and 950 nm. We performed multiple tests of read noise, dark noise, and linearity. The read noise is ~2 photons (RMS), the dark noise is 0.1 e- per second per pixel, and the response is linear within 0.3% (and perhaps better) over a dynamic range of a few to 90,000 photo-electrons. A fixed-pattern noise along columns is 1 to 2 electrons is removable with flat-field exposures. Also, "salt-and-pepper" noise results in +-10 electrons in a few hundred random pixels which are easily identified as they affect only single pixels. Importantly the CMOS sensor offers frame rates up to 48 frames/sec with a read-out time of ~20 ms, enabling sub-second exposures to improve the contrast betwesen light pulses having sub-second duration and the background "noise" of stars, galaxies, and sky.

From our two California observing stations at Anza Borrego and Taylor Mountain, the sky produces 20 and 45 photons/pixel, respectively, during 0.25 sec, causing a total 1-sigma noise of ~5 and ~7 photons/pixel per exposure. Each pixel is receiving sky photons from the full range of wavelengths, 360 to 1000 nm. We constructed a light baffle with a 5 deg opening angle, with its axis tilted 4 degrees to the telescope's optical axis to accommodate the 4.0 deg refractive bending of light caused by the prism.

Candidate pulses are confirmed or rejected using a second smaller objective prism telescope that operates simultaneously. The smaller telescope is an 80 mm aperture Stellarvue refractor with f-ratio f/5.25 and a ZWO ASI-1600 CMOS camera at its focus. An objective prism is mounted on the front, composed of the same glass with the same wedge angle as the primary Schmidt telescope. The ZWO ASI-1600 camera has a 4650x3520 format with 3.8 μm pixels, giving a similar field of view and spectral resolution as the larger primary telescope system. This camera is binned 2x2 and the system is 13x less sensitive than the larger Schmidt telescope, providing 3-sigma confirmation of threshold 10-sigma detections with the RASA. This refracting, objective prism telescope is mounted on the RASA telescope.

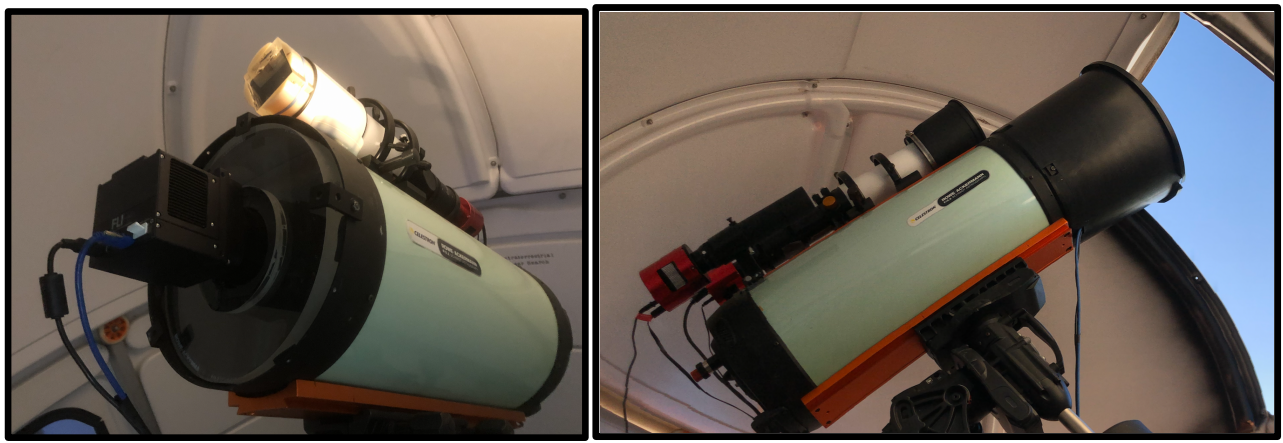

*Figure 2. The two objective prism telescopes. The 0.28-m RASA Schmidt is below, and the 0.08-m Stellarvue rides piggyback on top, ensuring the same field of view at all times. Left: The front baffles are off, allowing a view of the prisms. The FingerLakes KL400 CMOS camera (2kx2k, 11 μ pixels) is visible at the prime focus of the RASA. Right: The red ZWO ASI-1600 CMOS camera is visible (4656x3520, 3.8 μm pixels). Both cameras are thermoelectrically cooled to -20C. Both telescopes operate simultaneously with exposures of 0.25s to confirm or reject monochromatic, sub-second pulses of light.*

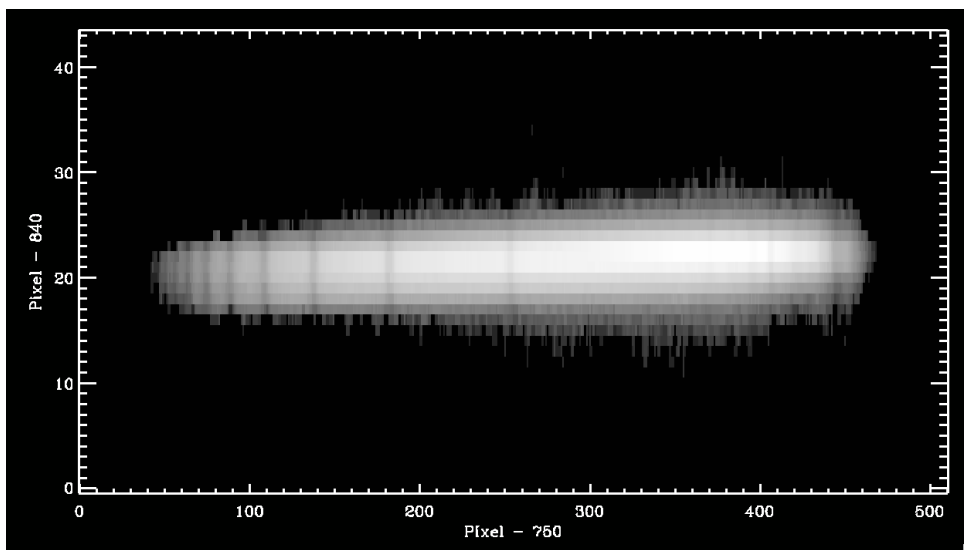

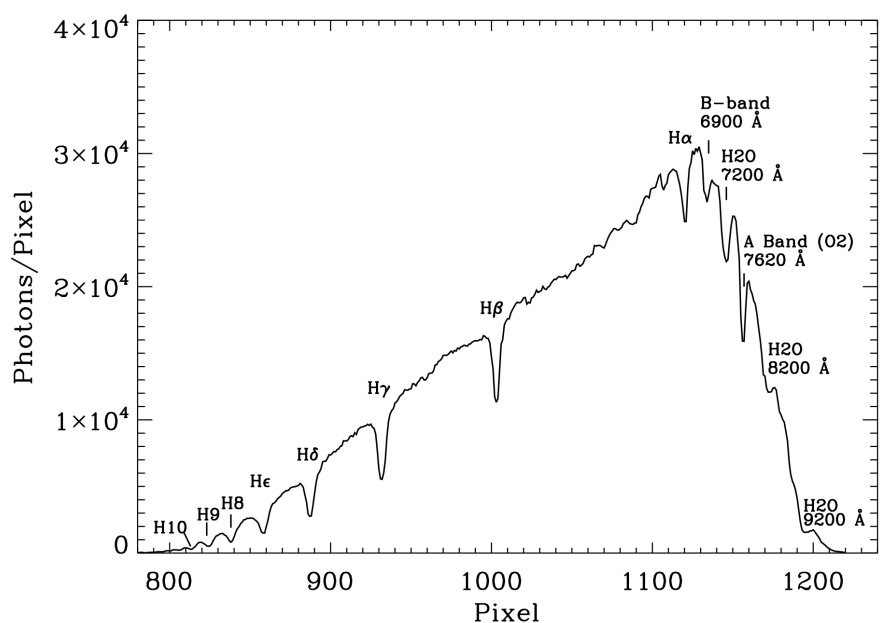

*Figure 3. A 12 ms exposure of Vega with the objective prism 0.28-m RASA Schmidt telescope. Top: Raw image using the KL400 CMOS camera, showing several absorption lines. Bottom: The extracted 1-D spectrum of Vega displayed as the number of photons (without flux calibration) versus pixel number toward increasing wavelength to the right along the Vega spectrum, with labelled absorption lines, notably the Balmer series and telluric lines. The spectrum spans 400 pixels, and the resulting spectral resolution varies from 10 to 25 Å from 3800 to 9500 Å, enabling the identification of narrowband (monochromatic) sources.*

We measured the wavelength scale and spectral resolution with a spectrum of Vega. Figure 3 shows a 25 ms exposure of Vega with the objective prism RASA Schmidt telescope. The raw image (Figure 3, top) shows Balmer and telluric lines. A simple reduction to a 1-D spectrum (bottom of Figure 3) was accomplished by summing the photons along the spatial width, revealing the absorption lines easily identified (as labelled in Figure 3) and exhibiting the nonlinear wavelength scale due to the higher refractive index of the prism glass at shorter wavelengths. The optical spectrum spans nearly 400 pixels, with the resolution set by the PSF with FWHM=2.5 pixels. The resulting wavelength calibration and the resolution of 2.5 pixels (PSF FWHM) show the spectral resolution varies from 10 to 25 Å between 3800 to 9500 Å, resolving the two [OIII] lines by 5.5 pixels as verified by spectra of planetary nebulae.

Figure 3 reveals the spectral response of the entire instrument system. The spectral energy distribution of Vega peaks near 400 nm and declines slowly towards longer wavelengths (roughly a Rayleigh-Jeans distribution). Figure 3, however, exhibits a peak near H-α at 656 nm in detected photon flux per pixel and a diminished response towards the blue and near-IR, falling to near zero at 380 and 950 nm. The primary contributor to the declining system response is the quantum efficiency as a function of wavelength of the Fingerlakes KL400 cmos camera, with secondary contributions from the coatings and glasses used in the RASA telescope which is optimized for the 400-700 nm wavelength region.

Exposures of photometric standard stars show that at magnitude V=10, the system yields 90 photons per pixel in a 0.25-sec exposure in the wavelength region 550-700nm spectrum. Noise from sky brightness is 7 photons (rms) per pixel, making spectra of magnitude V=12.5 stars just visible in our 0.25 s exposures.  Emission lines containing ~100 photons in the peak pixel will constitute ~10-sigma candidates.

### 3    OBSERVATIONS OF THE SOLAR GRAVITATIONAL LENS OF PROXIMA AND ALPHA CENTAURI

The SGL focal region of Proxima Centauri resides on a line that begins 550 au from the Sun at celestial coordinates opposite that of the star, RA = 2hr 29m 31.5, Dec = +62d 40' 30", for an observer located at the Sun at epoch 2021.0 and equinox 2000 (e.g., Figure 1 in Hippke 2021b).  Light rays from Proxima Centauri that pass by the Sun with larger impact parameter (farther from the Sun's surface) come to a focus farther than 550 au from the Sun, and the full set of rays creates a focal line.  In reverse, a transmitter could be located anywhere along that SGL line, beaming toward the Sun and bending the light onward to Proxima Centauri. For an observer on Earth, the apparent coordinates of such a transmitter must include corrections of up to 375 arcsec during a year caused by parallax to the transmitter (Hippke 2021b).  Relativistic aberration of light caused by the Earth's velocity induces an additional displacement of 20 arcsec.  The proper motion of Proxima Centauri itself amounts to 4 arcsec per year.

The intended recipient of the transmission is located at an unknown location near Proxima Centauri, leaving uncertain the most likely location of the local transmitter. Thus, during a year of observations, a transmitter located near the SGL focus line could appear within a region of roughly 400 arcsec from the coordinates given above. Parallax and aberration can be predicted for any instant, but the unknown location of the receiver near Proxima Centauri leaves the SGL location unknown within roughly 60 arcsec. We search within the full 400 arcsec domain for both sub-second pulses and long-lived emission of monochromatic sources.  Our total field of view of 2.2 x 2.2 deg easily includes that 400 arcsec domain.  Exploring the relatively large area around the anti-solar position of Proxima Cen expands the survey to include emitters located off-center that target only a fraction of the focal ring surrounding the Sun.

Similarly, the SGL focus of Alpha Centauri AB is centered at RA = 2hr 39m 19.4, Dec = +60d 49' 41", at epoch 2021.0 and equinox 2000. Any transmitter's instantaneous position near the SGL line must be similarly corrected by parallax, aberration, proper motion, and the unknown position of the intended receiver near Alpha Centauri AB.  Our observations of the SGL focal region include all RA and DEC within 400 arcsec of central position noted above and include the field of view within 1 degree of those coordinates, in both RA and DEC.

We obtained 88000 objective prism exposures of the SGL region of Proxima Centauri ( 6.1 hr total) and 47000 exposures of the SGL region of Alpha Centauri (3.2 hr total) during five months between 6 Nov 2020 and 10 Apr 2021. All exposures have a duration of 0.25 sec, with a negligible gap in time between exposures.  We only observe when the Moon is absent or less than the quarter phase, injecting a minor increase in the sky brightness of ~19 mags per square arcsec.

Figures 4 and 5 show a representative 0.25-second exposure of the SGL focal regions of Proxima Centauri and Alpha Centauri, respectively, taken with the objective prism 0.28-m RASA Schmidt telescope and the Fingerlakes CMOS KL400 camera, as described in Section 2.  The field of view is 2.2 x 2.2 deg, with north to the right and east up, so that the wavelengths within each spectrum increase from right to left. The stellar spectra span 400 pixels, covering wavelengths 380 to 950 nm (see Figure 3). The sky background is typically 45 photons/pixel (RMS ~ 7 photons/pixel), with each pixel covering 3.7x3.7 arcsec, and the faintest visible stars are magnitude 13.  Figures 4 and 5 include labels of several Hipparcos and Tycho stars for astrometric and photometric reference, all magnitude V = 8-10.  The dashed rectangles in Figures 4 and 5 enclose the approximate region of the SGL regions for Proxima and Alpha Centauri as seen from Earth, including parallax and aberration during a year.

s

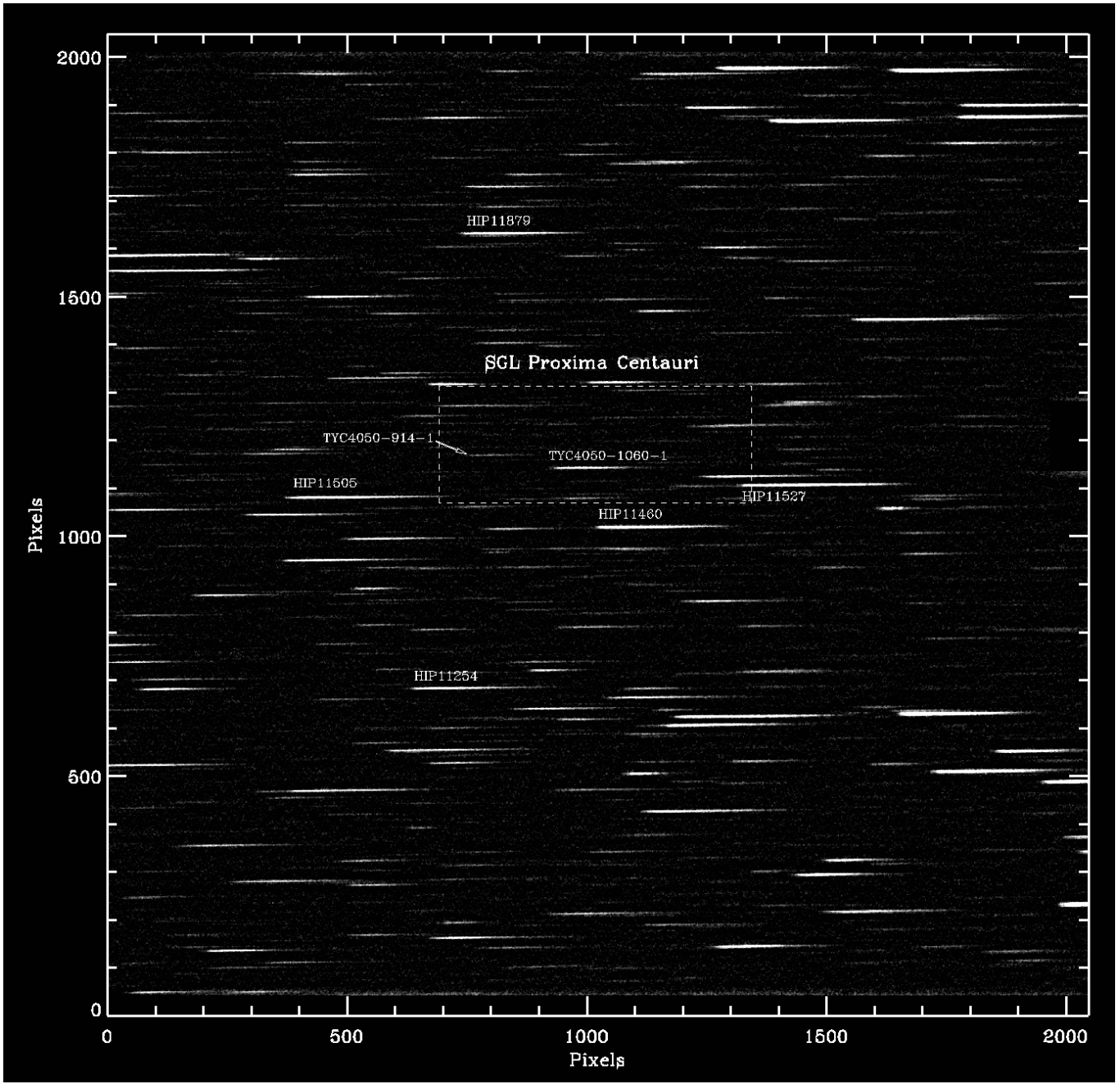

*Figure 4. A typical objective prism image with exposure time 0.25 sec centered near the solar gravitational lens focal region (dashed) of Proxima Centauri. The field of view is 2.2x2.2 deg, with north to the right and east up. The spectra span wavelengths 380 – 950 nm, with longer wavelengths to the left. Seven reference stars, Vmag = 8 to 12, are identified, and the faintest spectra come from stars having V ~ 13 mag. Monochromatic light would appear as a PSF-like dot.*

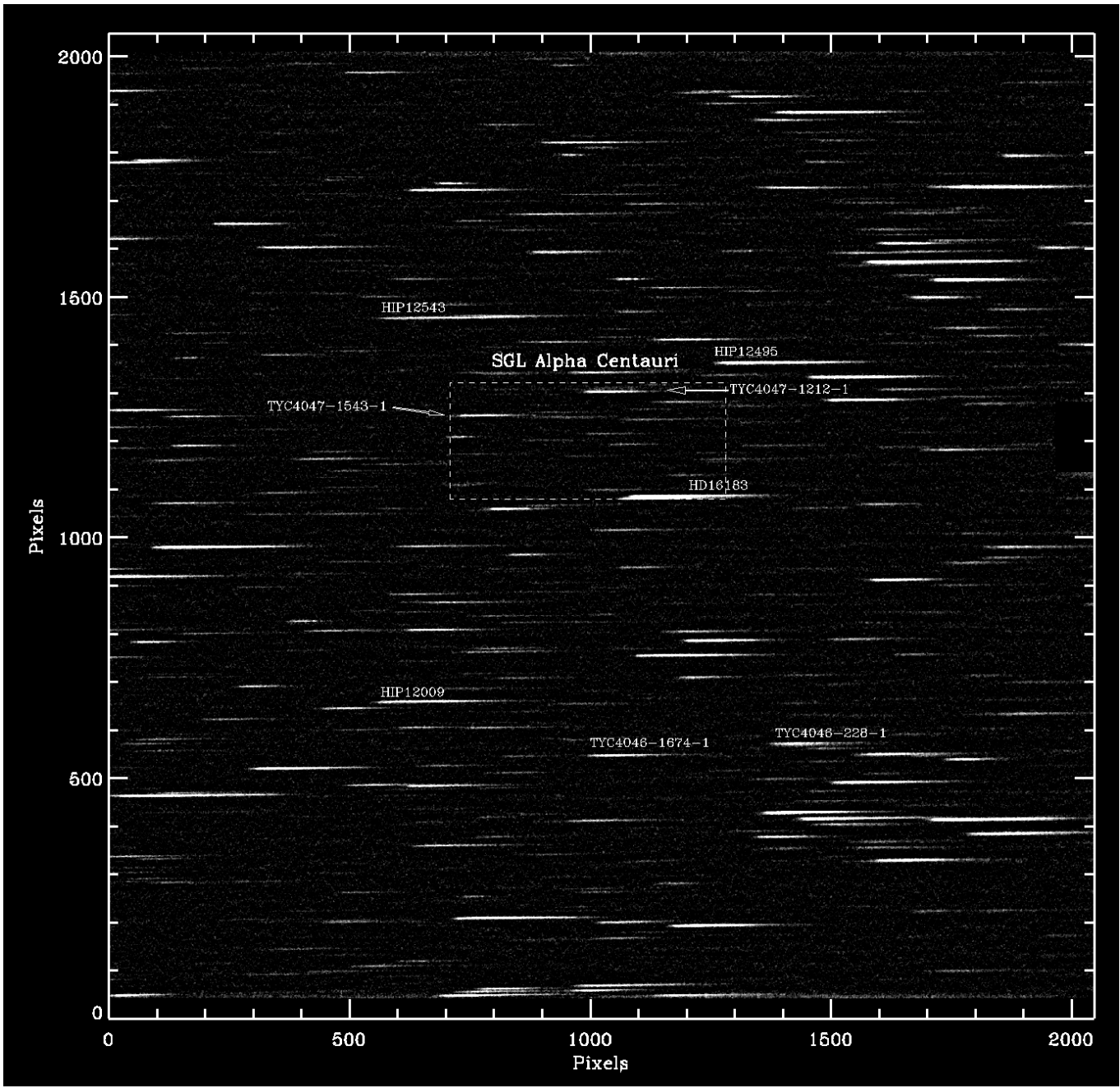

*Figure 5. Same as Figure 4 but for the SGL focus of Alpha Centauri AB. Eight stars, Vmag = 8 to 10, are identified for reference, and the faintest spectra have Vmag=13. Monochromatic sources would appear as dots with PSF shapes.*

The raw images in Figures 4 and 5 show the 0.25-sec exposures we use to search for pulses of monochromatic light from the 2.2x2.2 deg field of view, as described in Section 4. We verify any candidate pulses of light by using a second objective prism telescope described in Section 5 and provide the search results for pulses in Section 6. We also search for persistent "long-lived" monochromatic emission in the vicinity of the SGL focal regions of Proxima and Alpha Centauri. For that effort, we co-add 4000 consecutive images to reach detection thresholds of monochromatic emission 3.5 magnitudes fainter than possible with a single 0.25-sec image, described in Section 7.

### 4 THE SEARCH ALGORITHM FOR MONOCHROMATIC PULSES

We search for pulses of monochromatic emission by employing a difference-image technique to search for PSF-like "dots" that appear in a target image after subtracting the average of several images taken before and after. The emission dots may be coincident with stellar spectra or in the sky between them, and the subtraction suppressed both. Candidate dots must have a 2D shape consistent with the instantaneous point spread function (PSF), as exhibited in the spatial profile of the many stellar spectra. For every 0.25 s exposure, we measure the FWHM of the spatial profile, commonly 2.0 to 3.0 pixels, corresponding to 7.5 to 11 arcseconds, primarily due to optical aberrations in the prism.

The algorithm searches each "target" image in a sequence by subtracting the mean of six images composed of the three "bookend" images taken immediately before and after the target image. The resulting difference image has the persistent spectra of astronomical objects removed (mostly stars), including their emission lines. The subtraction is imperfect because of changes in intensity, image sharpness, and position on 10 millisec time scales caused by refraction within passing turbulence in the Earth's atmosphere. These "seeing" changes within 0.25 sec, along with simple Poisson noise, leave the residual differences as large as ~10% of the photons in the pixels in the original stellar spectra.

To mitigate the imperfect subtraction of the "bookend" images from the target image, we perform a boxcar smoothing along rows which are parallel to the dispersion in the spectra, using a boxcar width of 30 pixels. This captures the mean of the residual stellar spectra on scales of 30 pixels. (We also tried using median smoothing but settled with boxcar smoothing for its increased speed.) We then subtract that smoothed residual image from the original residual image, producing a new residual image having suppressed the stellar spectra further, leaving only the short-scale (<30 pixels) departures. Narrow emission lines with widths under 15 pixels remain undiminished in this new residual image.

We construct a metric of the positive departures from zero in the new residual image by computing the Poisson noise (from the original number of photons) and then computing the ratio of the new residual image to the noise to yield a local signal-to-noise ratio (SNR). Detectable emission lines will stand above unity in this ratio. We adopt a threshold of SNR > 4.5 and a minimum number of photons of 40 as the two criteria that define individual "pixels of interest." By searching all rows, we obtain a list of all individual pixels meeting those two criteria, without regard (yet) to the 2D profile of the candidate emission.

We examine the pixels near these individual pixels of interest to determine if the counts in them are consistent with the 2D point-spread function (PSF). We use the spatial profile of the stellar spectra as a direct measure of the PSF, and we assume it is nearly symmetrical in both the spatial and dispersion directions. Indeed, H-alpha emission from flare stars has the same PSF in the dispersion direction as the spatial. The prism and telescope should cause little or no anamorphic distortion.

We compute the RMS of the difference between the candidate "dot" and the measured, instantaneous PSF (from the spatial profile) from that same image, both being normalized to the peak. We retain all dots having a PSF shape with RMS less than 10% of the peak as viable monochromatic pulse candidates.

## 4.1 Injection and Recovery of laser pulses

To test the algorithm and determine detection thresholds, we generated 25 synthetic monochromatic pulses consisting of 2D Gaussians having FWHM=2.5 pixels similar to the typical PSF. We scaled these pulses to various counts in the peak pixel, from 100 to 500 counts, representing 2x to 10x the background sky counts per pixel. We added these synthetic pulses to actual individual images (implying a pulse duration less than 0.25 s), and we placed the pulses at locations both in between and coincident with stellar spectra. Blindly executing the difference-image algorithm described above, we found the code successfully "discovered" 100% of the injected pulses that had at least 55 photons in the peak pixel, but it found only 50% of the injected pulses that had 40 photons in the peak pixel.

Figure 6 shows four representative synthetic laser pulses injected into a raw image. The laser pulses have 55 to 110 photons in the peak pixel of the PSF, and all have the shape of a 2D Gaussian with an FWHM of 2.5 pixels. The code discovered all of them, no matter where they were located within the 2.2 x 2.2 deg field of view. Injecting weaker synthetic pulses showed that pulses with only 40 photons in the peak pixel were detected in only 50% of the trials. Such pulses contain 200 photons total within the 2D profile of the PSF. Thus, the detection threshold at which 50% of the pulses would be detected is 200 photons total within the pulse. *This 200 photon threshold represents a characteristic detection threshold photon fluence in a 0.25 s exposure for which half of the pulses would be detected.* Pulses with 300 photons total in the PSF are almost all detected, as they sit well above the sky noise.

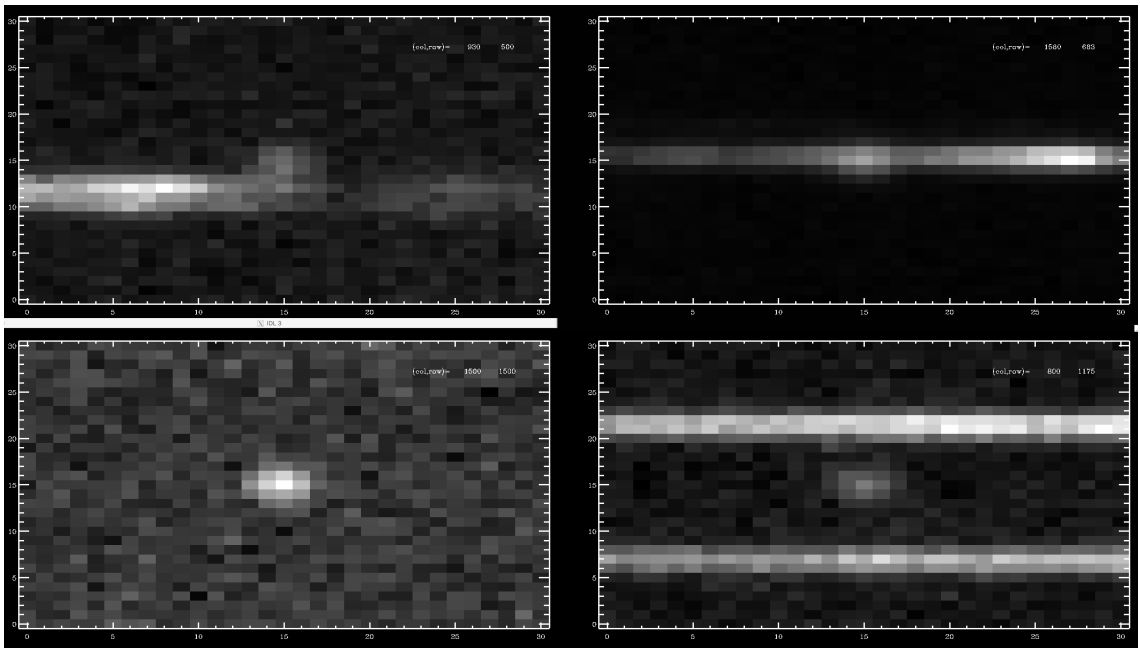

*Figure 6. Images of four synthetic monochromatic emission pulses representative of the 25 PSF-like pulses we added to actual observed 2.2 x 2.2 deg images as a test of the efficacy and detection threshold of the algorithm. The synthetic pulses contained between 100 and 200 photons in the peak pixel. This is more intense than the nominal algorithm threshold of 40 photons in the peak pixel, at which detection efficiency falls to 50%. Some synthetic dots are coincident with stellar spectra and others not. The search algorithm blindly "discovered" all four dots and identified no others in the images, representing efficacy and low false-positive rate for PSF-like pulses.*

Monochromatic pulses of light with durations much shorter than 0.25 sec are detectable, including durations as short as a nanosecond. The only requirement is that the total number of photons (fluence) during a 0.25 sec interval is above the threshold of 200 photons in the pulse. The light must be "monochromatic" within the spectral resolution, varying from 10 to 25 Å from 3800 to 9500 Å. The most powerful MegaWatt and GigaWatt pulsed lasers meet this monochromaticity criterion.

This search algorithm has diminishing sensitivity for pulses lasting 0.25 to 1 sec for which some of the six bookend exposures would contain the emission, diminishing their contrast with the target image. To detect light pulses longer than 1 sec, we employ a different algorithm described in Section 7.

## 5    CONFIRMING PULSES WITH AN OBJECTIVE PRISM REFRACTOR

We confirm or reject emission pulses from the RASA telescope by using a second, smaller objective prism telescope operating simultaneously, as described in Section 2. This smaller telescope acquires images with the same cadence of 4 frames per sec and is centred at the same RA and Dec on the sky. It has a slightly different aspect of 1.8 x 2.5 deg than the RASA 2.2x2.2 deg, thus giving incomplete overlap of the image from both telescopes. However, the two SGL regions with only 400 arcsec across enjoy 100% overlap between the RASA and Stellarvue telescope, thus enabling confirmation of both pulsed and persistent lasers.

We measured the relative photon-collecting efficiency of the two telescope systems by taking 100 exposures of 0.25 sec duration with both telescopes pointed at the same star field. We identified one $8^{th}$ magnitude star for the study. After extracting the spectra and normalizing the continua to unity, we subtracted adjacent exposures and computed the difference in each of 200 pixels along the spectrum. The differences are positive and negative, resulting from Poisson noise in the photons arriving at each wavelength. The RMS of the fractional difference among the 100 spectra is a direct measure of the fractional noise, as the stellar spectra do not change measurably within 0.25 sec.

These RMS values indicate the ratio of photons collected is 13 between the two telescopes assuming Poisson fluctuations of photon arrival dominate. Indeed, the ratio of aperture areas of the two telescopes is $11^2/3^2=13$. The 30% central obstruction of the RASA telescope and QE of the two sensors, 80% and 60%, nearly cancel their effect on the relative throughput. Thus, the ratio of detected photons and apertures is in rough agreement at 13. A monochromatic pulse in the RASA system that contains a threshold of 200 photons within the 2D PSF will yield 15 photons within the 2D PSF detected by the Stellarvue system. The sky brightness in Stellarvue images is typically 3 photons per pixel, leading to 15 photons within a 2D PSF and Poisson fluctuations of ~4 photons (rms). Thus, the threshold of 15 photons in the pulse is ~3.5 $\sigma$ above the noise in the 2D PSF, allowing the small 80-mm Stellarvue telescope to confirm or reject a 200 photon emission line found by the RASA telescope, a threshold event.

## 6  RESULTS: THE SEARCH FOR MONOCHROMATIC PULSES

Using the double objective prism system described in Sections 3, 4 and 5, we searched for sub-second pulses of monochromatic light in each of the 88000 exposures of Proxima Centauri and the 47000 exposures Alpha Centauri, all exposure times being 0.25 s. Using the larger RASA telescope alone, we found 58 candidate monochromatic light pulses from the SGL region of Proxima Centauri and 19 candidate monochromatic pulses from the SGL region of Alpha Centauri. For each candidate monochromatic pulse found in a RASA image, the smaller Stellarvue telescope provided two exposures within 0.25 s, one before and one after. The two telescopes were not synchronized, but they both maintained a 0.25 sec cadence and exposure time.

To confirm or reject each candidate monochromatic pulse found in a RASA image, we examined by eye the two Stellarvue exposures time-centered within 0.25 s of it. One Stellarvue image was always centered within 0.125 sec of the RASA image. We used stellar spectra located within a few arcminutes of the candidate monochromatic dot as astrometric references. This astrometric approach has uncertainties of ~5 arcsec due to the width of the PSF and the length of the spectra, but it is adequate to predict where a dot should appear in the Stellarvue image with an uncertainty of ~2 pixels (7 arcsec).

Figure 7 shows four representative candidate monochromatic pulses found in RASA images. The left image of each of the four panels shows the PSF-like dot found in the RASA image. The right image shows the nearly simultaneous Stellarvue image. The monochromatic "dot" in the RASA image does not appear in the Stellarvue image for each left-right pair of images. In searching for the corresponding "dot" in the Stellarvue image, one must account for the factor of 13 lower sensitivity of the Stellarvue by using neighboring stars as photometric references.

Figure 7 shows that none of the four representative monochromatic pulses in the RASA images also appears in the contemporaneous Stellarvue image. If the pulses were coming from actual point sources in the sky, the dots in the RASA images would have appeared in the Stellarvue images after accounting for the factor of 13 lower throughput. None of the 77 monochromatic pulses seen as "dots" in RASA images appear in the Stellarvue images. Their absence forces us to reject all of them as candidate monochromatic emissions. *We found no sub-second pulses of monochromatic light from the SGL regions of Alpha or Proxima Centauri within 2.2 x 2.2 deg.*

We suspect the PSF-like dots in the RASA images are caused by elementary particles directly hitting the CMOS sensor. Tests of the Fingerlakes KL400 CMOS camera involving 10 min dark exposures reveal that elementary particles usually deposit electrons in only one or two pixels. Occasionally they deposit electrons in 4 to 8 adjacent pixels, with one or two central pixels containing more electrons than the periphery, not unlike the shape of the PSF. Such a particle hit occurs in the CMOS sensor roughly every half hour, depositing electrons into neighboring pixels in a pattern that masquerades as PSF-like monochromatic pulses. Indeed, such direct hits of particles are why we designed and installed the Stellarvue objective prism system. We do not know if the elementary particles are cosmic rays, solar wind particles, or natural radioactivity from the Earth or materials, nor do we know what types of particles predominate. We occasionally see a cluster of particle hits on the CMOS sensor, indicating that a particle shower was launched nearby.

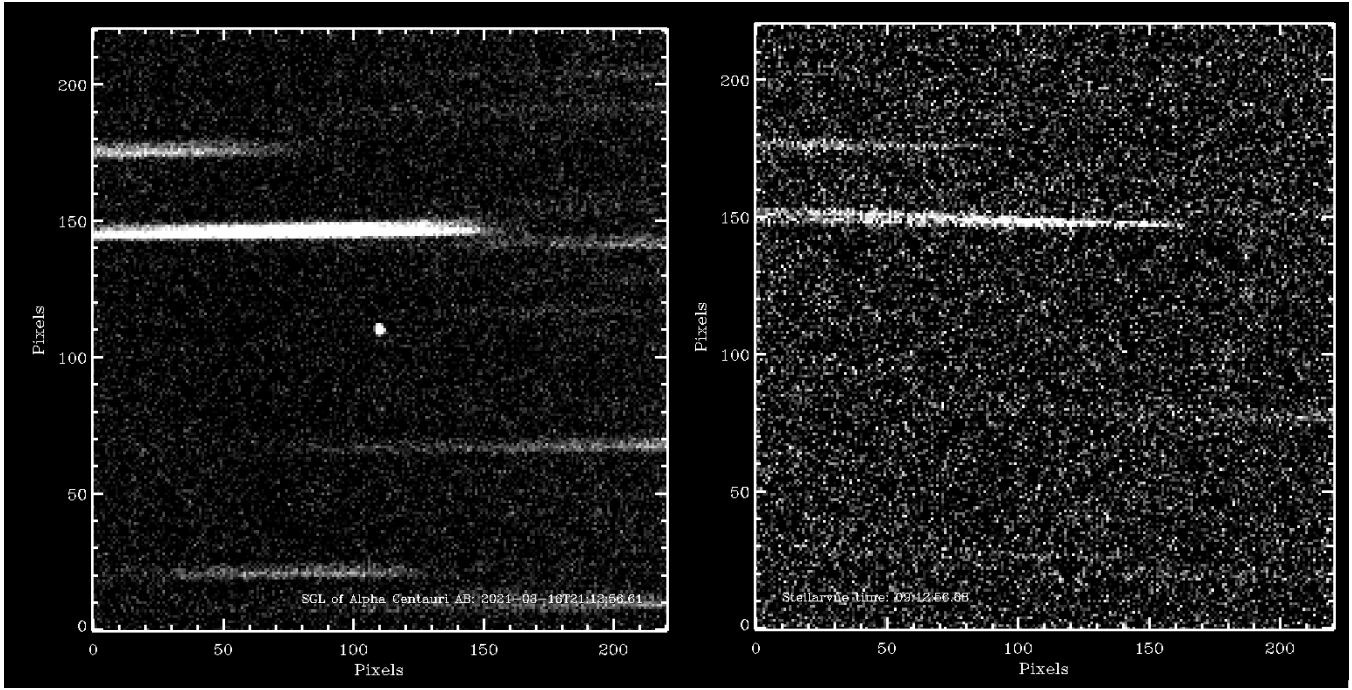

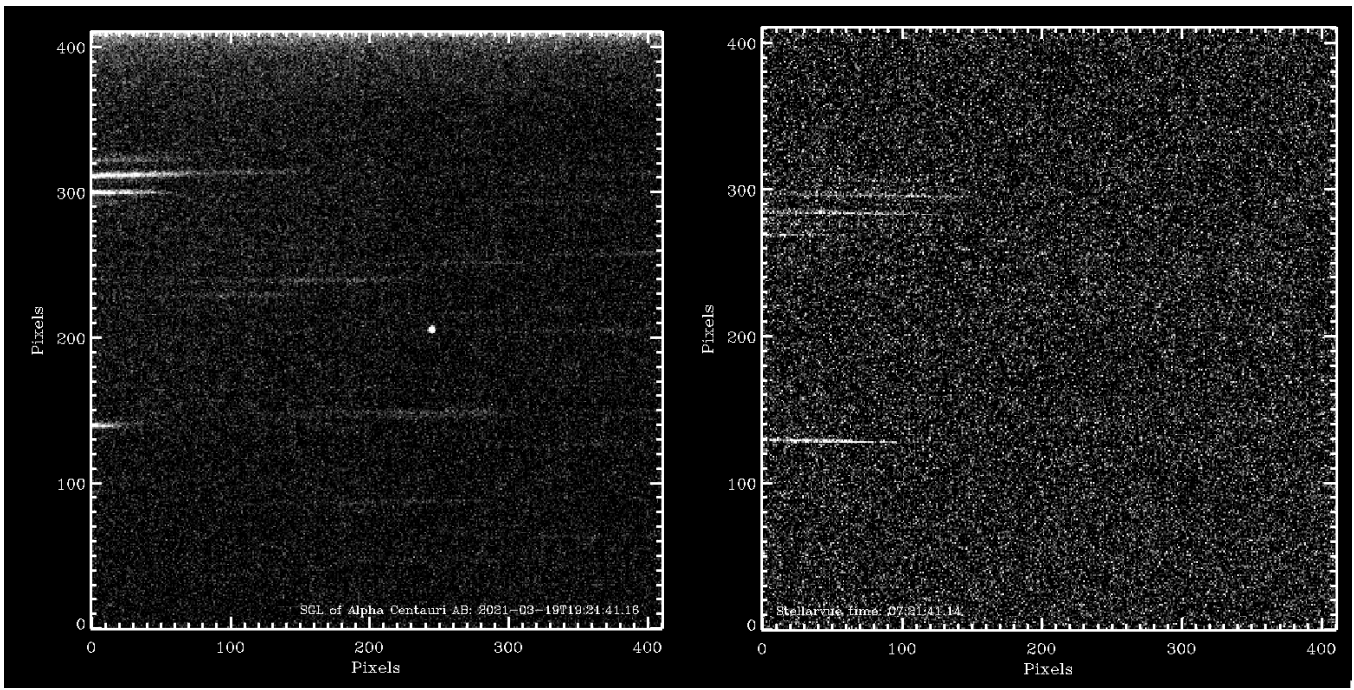

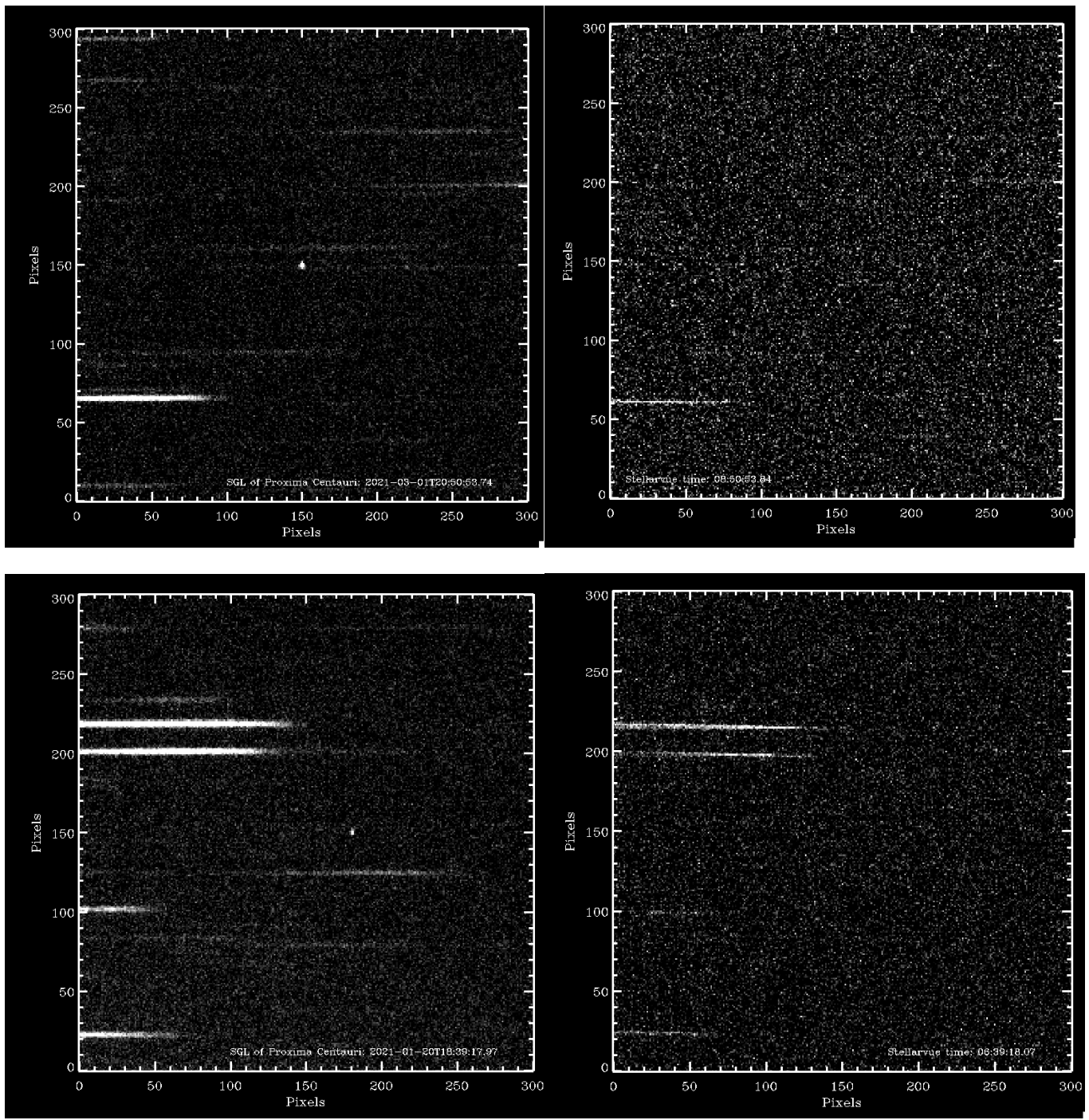

*Figure 7: Four candidate monochromatic flashes observed with the RASA telescope (at left), checked with the exposures (at right) made within 0.125 s with the Stellarvue telescope. All exposure durations are 0.25 s. Top pair: a PSF-like dot near the SGL of Alpha Centauri AB on 17 Mar 2021 at 4:12:56.61 (UT). The Stellarvue exposure (at top-right) was centered 0.07 s later, but the candidate "dot" does not appear, rejecting it as caused by light. $2^{nd}$ row pair: near the SGL of Alpha Centauri AB on 20 Mar 2021 at 2:21:41.16 (UT). The Stellarvue exposure was centered 0.02 s later, but the candidate "dot" does not appear, rejecting it. Third pair: near the SGL of Proxima Centauri on 2 Mar 2021 at 04:50:53.74. The candidate monochromatic pulse does not show up in the simultaneous Stellarvue image. Bottom pair: near the SGL focus of Proxima Centauri on 21 Jan 2021 at 2:39:17.82 (UT). The candidate pulse does not appear in the simultaneous Stellarvue image. These four pairs are representative of the 77 cases where the PSF-like emission in the RASA telescope was absent in the Stellarvue telescope, rejecting the candidate.*

# 7 THE SEARCH FOR LONG-LIVED MONOCHROMATIC EMISSION

We also searched for long-lived monochromatic optical emission coming from near the SGL focal regions having a duration longer than 1 second and thus not detectable in the "difference-image" approach described in Section 4. We co-added sequences of 4000 images to form summed images having total exposure times of 1000 s. We performed such co-additions among the 88000 exposures of the SGL focus region of Proxima Centauri and the 47000 exposures for Alpha Centauri, all obtained during the 6-month observing season described in Section 3.

During 1000 seconds, our unguided telescope drifts by 30 to 60 arcsec in RA and by a few arcsec in DEC, necessitating registration before co-adding the images. We determine the relative displacements of successive images using a simple chi-square minimization (essentially a cross-correlation) in the RA direction with a precision of ~0.1 pixel. We identify a "template" image in the middle of the image sequence for the chi-square minimization. We do not determine the displacement in DEC as it is less than our Nyquist spatial resolution of ~7 arcsec. Each image is then shifted using a simple linear interpolation to accommodate displacements of fractional pixel amounts before adding it to the other images. This co-addition of 4000 images causes an additional blurring of ~0.2 pixel, resulting in a small resolution loss compared to the PSF FWHM of 2.5 pixels.

The resulting SGL focal regions are shown in Figures 8 and 9 with coordinates relative to the SGL focus as seen from the Sun. The grayscale is adopted so that the faintest stars have SNR ~ 5 per pixel at Vmag = 14.5. The noise comes from Poisson fluctuations of the photons and sky brightness fluctuations. The brightest stars in Figures 8 and 9 are Vmag = 9, while the faintest visible have Vmag=14.5. Any persistent monochromatic emission would be preserved and would appear as a PSF-like dot. We searched all 88000 and 47000 exposures within co-added sequences of 4000 exposures of the SGL regions of Proxima and Alpha Centauri. No PSF-like dots were found by eye in these co-added images.

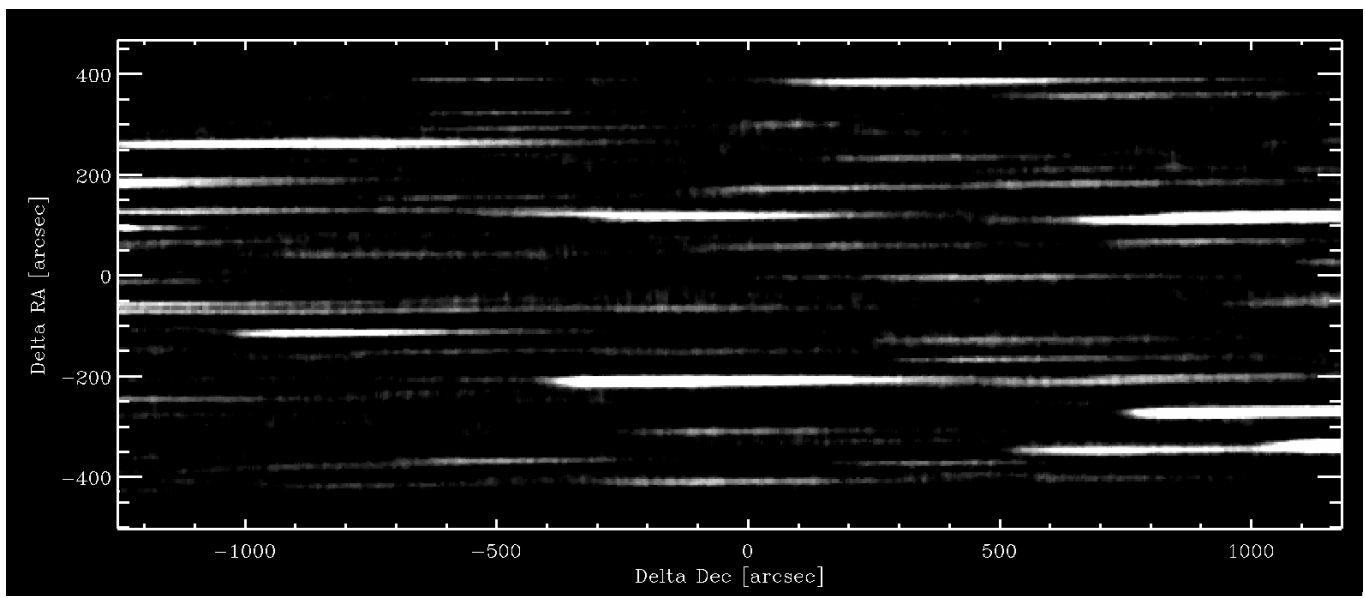

Figure 8. A typical co-added image, 16.7 min total, of the SGL focal region for Proxima Centauri. The axes represent angular distance in arcseconds from the solar gravitational focal point of Proxima Centauri for an observer at the Sun. Parallax and aberration from the Earth's orbital motion displace the apparent position of the SGL by up to 400 arcsec during the year, contained within this rectangular field of view. The horizontal streaks are spectra of stars spanning the magnitude range, Vmag = 9.8 to 14.5. The Declination coordinate is registered at 5000 Å in the spectra. The bright star 200 arcsec below center is TYC4050-1060 (Vmag=9.9). Any PSF-like dot would represent monochromatic emission. None appears, as discussed in Section 6.

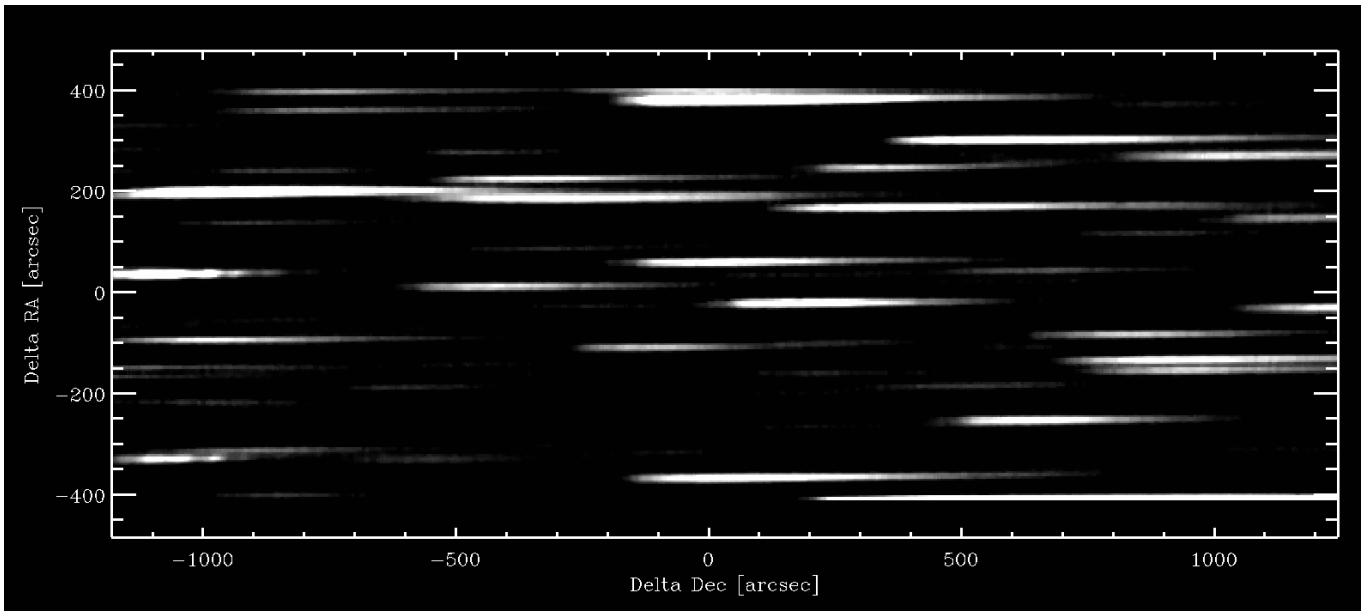

*Figure 9. A typical SGL focal region for Alpha Centauri in a 16.7 min exposure (4000 x 0.25-sec exposures), similar to Figure 7. Parallax and aberration displace the apparent position of the SGL by < 400 arcsec, contained within this rectangular field of view. Any PSF-like dot would represent monochromatic emission. None appears.*

To probe as deeply as possible, we rescaled the co-added exposures. Figure 10 shows a representative co-added exposure (4000 x 0.25s) for Alpha Centauri displayed with a grayscale tuned to reveal the faintest detectable PSF-like monochromatic emission visually. The two axes represent angular distance in the sky, with the origin being the SGL focal point as seen from the Sun. We examined only the region within +-450 arcsec in DEC and RA from that sub-solar SGL focal point to encompass displacements due to parallax and aberration as viewed from Earth. The entire region examined for monochromatic emission is shown in Figure 10. The extent in RA is 1200 arcsec from the center of the SGL focal region to accommodate the spectral dispersion in that direction. Emission from 3800 to 9900 Å would be detected.

Figure 10 shows several dozen horizontal streaks that are the spectra of background stars having magnitudes Vmag = 9 to 15. The co-added exposures for both Alpha Centauri and Proxima Centauri contain similar densities and brightnesses of the background stars as displayed in Figures 8 and 9, yielding similar detection thresholds. We performed 20 trials with synthetic emission lines added to the individual images prior to co-addition to gauge the approximate detection threshold of monochromatic emission.

At faint flux levels similar to ~15$^{th}$ mag stars, we see randomly placed "dots" appear throughout the co-added images. The dots are caused by the spectra of faint background stars, Vmag=15 – 17, and the many gaps between their absorption lines (stellar and telluric) scattered throughout the image. The resulting intensity variations above and below that background intensity produce "dots" that masquerade as monochromatic emission, just barely visible in Figure 10. We set the preliminary detection threshold for monochromatic emission a factor of 2 above that background stellar variation, corresponding to the continuum flux density of approximately a Vmag ~ 14 star. This preliminary threshold corresponds to the fluence of Vmag=14 star within the PSF with a FWHM given by $\lambda/\Delta\lambda = 100$.

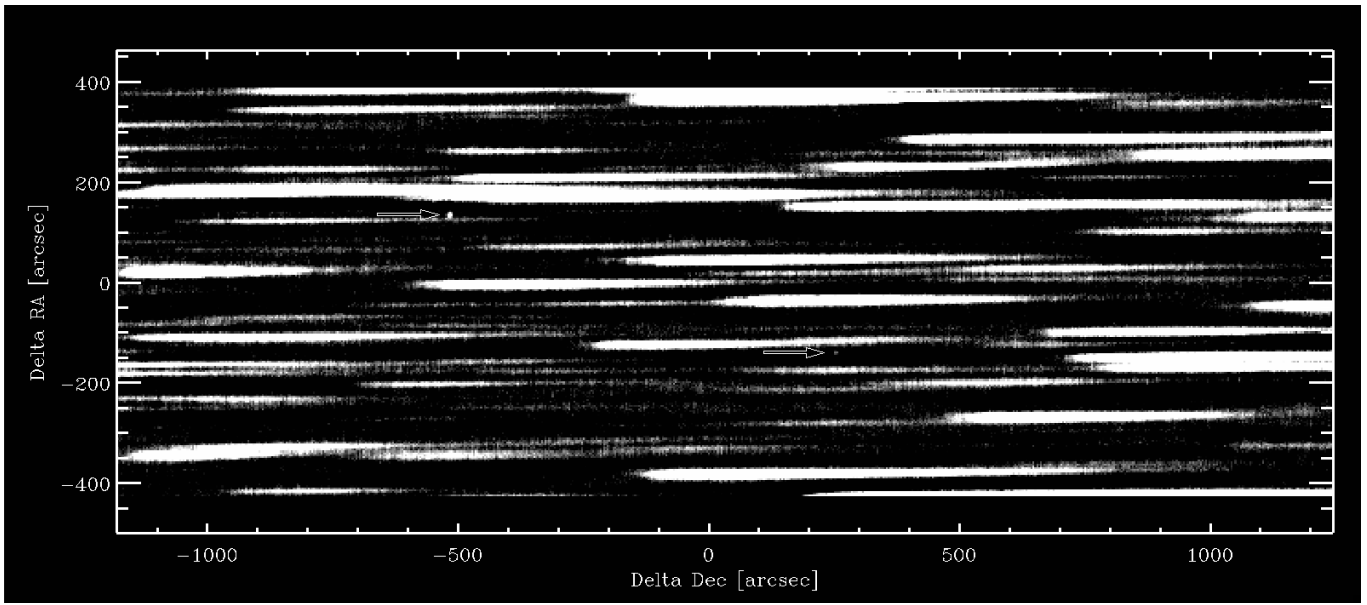

*Figure 10. The SGL focal region for Alpha Centauri in a 1000 s exposure (4000 x 0.25 s, as in Figure 8) but h two synthetic monochromatic emission lines (see two arrows). The synthetic emission has a PSF-like shape – a 2D Gaussian with FWHM=2.5 pixels. Here, the grayscale is tuned to optimize the search for monochromatic emission detectable in a 1000 s integration time. The synthetic "dot" at upper left consists of 3 photons (in the peak pixel of the PSF) per exposure and is clearly detectable. The dot at the lower right has only one photon per exposure and would not be securely detected as many false positives would be indistinguishable. The threshold for detection is thus ~2 photons per 0.25 s exposure in the peak pixel of a monochromatic emission line.*

Figure 10 shows two synthetic, monochromatic emission sources, appearing as dots, constructed with simple 2D Gaussians to simulate the PSF. They have 1 and 3 photons in the PSF peak and are injected each 0.25 s exposure. In the co-added sum, the synthetic emission dot constructed with 3 photons in the peak is clearly visible, while that with 1 photon in the peak is only marginally detectable. The preliminary detection threshold thus corresponds to the monochromatic emission of 2 photons at the peak of the PSF-like emission in each 0.25 s exposure.

*Examination of all of the (co-added) 1000 s exposures revealed no monochromatic emission within the SGL focus regions of Alpha Centauri and Proxima Centauri.* We would have reported any candidate monochromatic emission that might merit reobservation, but no such candidates emerged. We thus find no evidence of emission above our detection threshold, near the continuum flux of a star of brightness, Vmag=14, sampled at our spectral resolution, $\lambda/\Delta\lambda \sim 150$, i.e., a resolution of ~4 nm per 2.5 pixels at 600 nm.

We note that our bandpass is 1/25 of typical optical broadband photometry that is ~100 nm. Thus, in past broadband photometric surveys, our threshold monochromatic emission would appear roughly 25x fainter, namely 17.5[th] magnitude, because the monochromatic emission fills only 1/25 of the broad wavelength bandpass. In particular, at 600 nm, our spectral resolution is 4 nm instead of 100 nm in broadband imaging. Monochromatic light at threshold intensity here and emitted continuously would appear magnitude ~17.5 in a broadband survey. Such emission-line objects would be both detectable and notable here because of the immediately apparent monochromatic emission.

We more accurately measured the detection threshold for long-lived monochromatic emission by performing injection-and-recovery tests. We modelled monochromatic emission as PSF-like dots computed as 2D Gaussians with FWHM = 2.5 pixels and added them at random pixel locations to images that were actually observed. We injected such synthetic monochromatic emission, with a wide variety of intensities, into each of 4000 consecutive images that were actually observed. We shifted and co-added the images, as described above in our standard algorithm. We searched the resulting co-added image for any monochromatic emission, using the same grayscale and threshold as the actual search.

We note that monochromatic emission could appear coincident with a background star. We thus designed and tested the detection algorithm for monochromatic emission lines competing against just background sky brightness and also against background stellar spectra. Any emitter aligned with a background star will suffer from additional competition from the variations in the stellar spectrum with wavelength.

We enhanced the detectability of long-lived monochromatic emission by taking the co-added image and performing a 41-pixel median smoothing in the horizontal direction. This has the effect of smoothing the stellar spectra, 400 pixels long, along the direction of wavelength dispersion. We subtracted that median-smoothed image from the original co-added image to yield an image having the stellar spectra mostly removed, except for strong absorption and any emission lines. The strong telluric lines in the red and near-IR leave behind a string of four residual dots to be ignored here.

The Proxima and Alpha Centauri results are shown in Figures 11 and 12, respectively, along with synthetic PSF-like monochromatic emission corresponding to 1 and 5 photons at the peak of the PSF. The resulting detection threshold remains around 2 photons per 0.25 s exposure in the peak of the PSF-like emission.

Figure 11 shows the co-added image of Proxima Centauri, with two synthetic point sources located at columns 1260 and 1280 and row 860. This is a typical 16.7 min exposure (4000x0.25 s), as in Figures 9, 10, and 11. The synthetic PSF-like emission coincides with the spectrum of a Vmag=13 star, making detection of monochromatic emission more difficult. The top panel of Figure 10 is the co-added 16.7 min exposure, with two injected synthetic monochromatic sources. The bottom panel is the same image after subtracting a 41-pixel median smoothing in the horizontal direction to remove the continua of stellar spectra. The synthetic source on the right (5 photons at PSF peak each 0.25 s) is easily detectable. The source with only 1 photon injected each 0.25 s is only marginally detectable. Thus, the detection threshold is ~2 photons per 0.25 s (peak) per exposure for such a 16.7 min co-added exposure, as we had found before. For comparison, the sky brightness contributes 25 photons per pixel in 0.25 s. Of course, background subtraction reduces the sky brightness, except for fluctuations. Figure 12 is similar to Figure 11, but for Alpha Centauri.

We found that synthetic monochromatic emission having 2 photons in the PSF peak pixel in an individual image was clearly detectable in the co-added image, as shown in Figures 8, 9, and 10. Synthetic emission having only 1 photon in the peak pixel co-added to a final synthetic "dot" only added up to a marginal detection, as shown in the lower right of Figure 10 and in Figures 11 and 12. We find the detection threshold is ~2 photons for the peak pixel of monochromatic emission in each 1000 individual 0.25 s exposures.

Integrating over the 2D PSF (i.e., in both RA and DEC directions), the detection threshold for long-lived monochromatic emission is ~15 photons per 0.25 s exposure. This threshold of 2 photons in the peak, and 15 photons within the entire PSF, is consistent with the local night sky brightness that delivers 25 photons per 0.25 s exposure per pixel, causing Poisson fluctuations from pixel to pixel of ~5 photons. *In summary, any monochromatic light source from the SGL focus regions of Alpha Centauri or Proxima Centauri that delivered at least 15 photons per exposure would have been detected. No such emission was found.*

This threshold of 15 photons detectable by the telescope in 0.25 s can be translated into a flux threshold at the Earth by using the aperture 0.27-meter of the telescope and the efficiency of the optical system. The RASA telescope collecting area is 0.040 m², and the system throughput is 0.49, including all optics and the ~80% QE of the CMOS sensor. This yields an effective collecting area of 0.020 m².

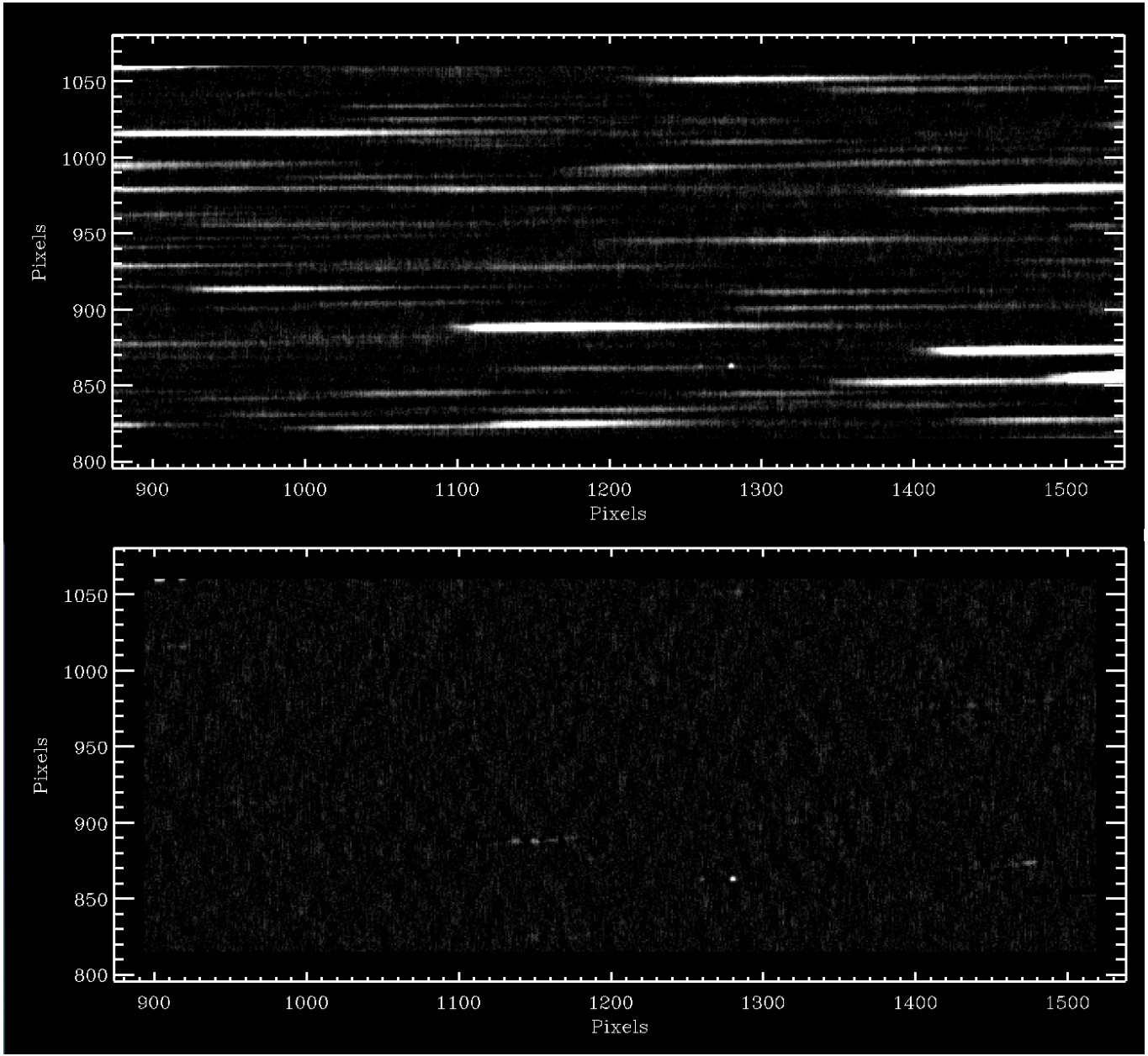

Figure 11. The SGL focus region of Proxima Centauri with two synthetic monochromatic point sources was injected containing 1 and 5 photons at the peak of the PSF into each 0.25 sec exposure. The two sources are located in columns 1260 and 1280 and row 860. This is a typical 16.7 min exposure (4000x0.25 s), as in Figure 7. The synthetic emission is coincident with the spectrum of a Vmag=13 star. Top: the 16.7 min exposure, with two injected synthetic monochromatic sources. Bottom: The same image after subtracting a 41-pixel median smoothing to remove the stellar spectra. The synthetic source on the right made with 5 photons per 0.25 sec is easily detectable, while that with 1 photon per 0.25 sec is marginally detectable. The detection threshold is ~2 photons per 0.25 sec (peak).

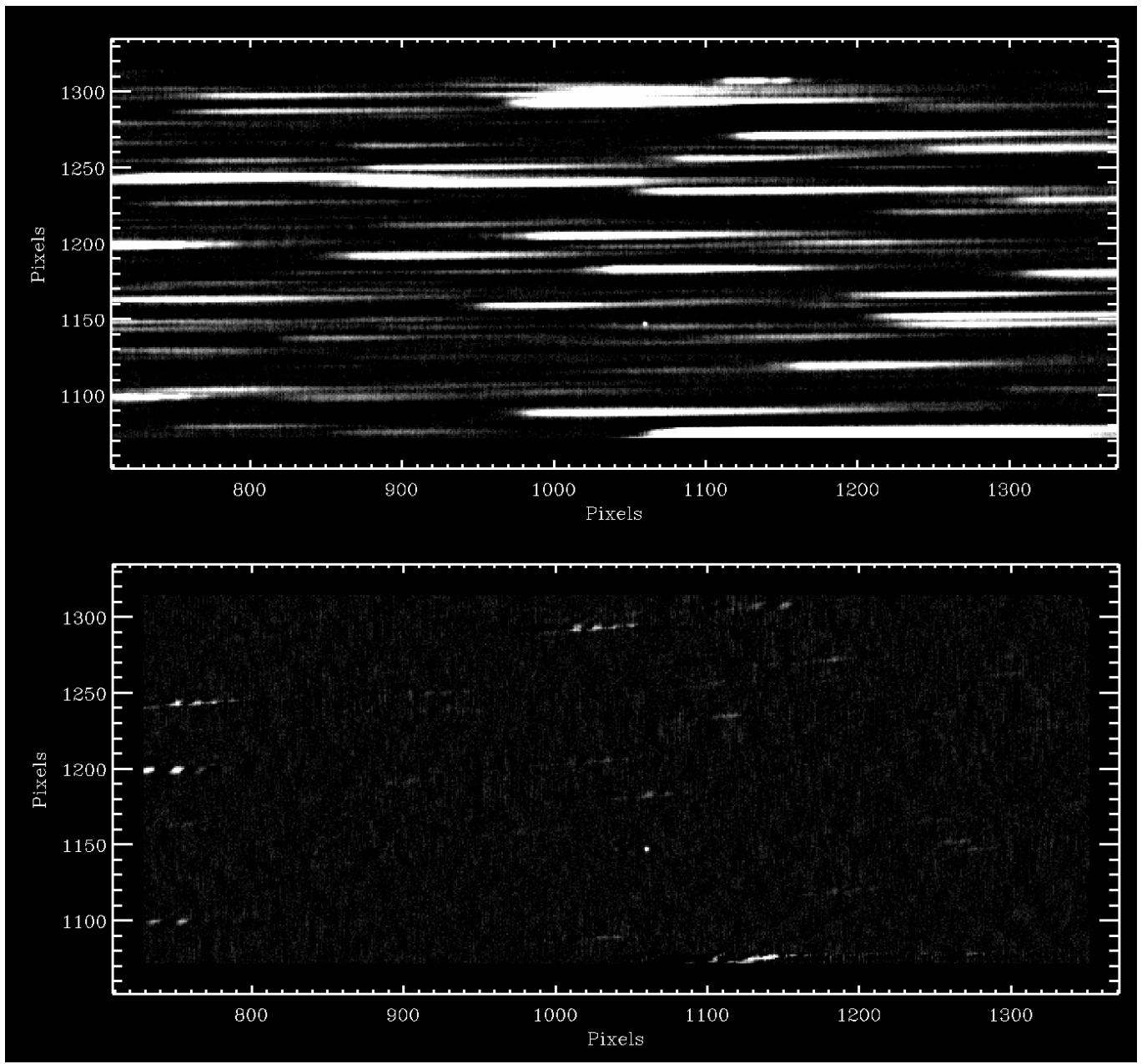

*Figure 12. The SGL focus region of Alpha Centauri similar to Figure 10. Two synthetic monochromatic point sources were injected, having 1 and 5 photons at the peak of the PSF in each 0.25 sec exposure, located at columns 1040 and 1060 and row 1150. Top: the 16.7 min exposure, with two injected synthetic monochromatic sources. Bottom: the same image after subtracting a 41-pixel median smoothing to remove the stellar spectra. The synthetic source on the right containing 5 photons (peak) per 0.25 sec is easily detectable, while the source made with only 1 photon per 0.25 sec (peak) is marginally detectable. The detection threshold is ~2 photons (peak pixel) per 0.25 sec exposure.*

Thus the flux threshold for long-lived monochromatic emission is 15 photons per 0.25 s per 0.02 $m^2$ = 3000 photons per $m^2$ per sec. Any source of monochromatic optical light in the SGL focus regions of Proxima or Alpha Centauri delivering 3000 photons per $m^2$ per sec at the Earth would have been detected. No such emission was found.

## 8 PROBE STATIONKEEPING AND POWER REQUIREMENTS

An SGL probe must remain aligned with the Sun and the transceiver at Alpha Centauri, as shown in Figure 1. The required positional accuracy depends on the size of the point spread function (PSF) of the gravitationally focused light (Hippke 2021a). Any errors in the probe's position by more than the size of the PSF will result in loss of light, which compromises the gravitational lens strategy. The PSF may be as small as ~0.1 meter in size for optical light, and its size depends inversely on wavelength (Hippke 2021a). Optical engineering challenges may result in a larger PSF, and the size of the telescope aperture must be as large as the PSF to capture most of the light. However, the probe's telescope will likely remain as small as possible to minimize payload. One may consider a benchmark PSF and telescope that are 1 meter in size.

The probe must position itself at the gravitational lens focus within a tolerance of that PSF size and telescope aperture. However, the focus is a moving target. The Sun is not in an inertial frame but instead accelerates due to the gravitational pull on it by Jupiter, Saturn, and the other planets. The Sun's resulting curlicue motion displaces it by 1.2 million km from the barycenter of the Solar System during a decade. The probe must slave, within ~1 meter, to that 1.2 million km motion of the Sun to remain aligned with the transceiver at Alpha Centauri.

Similarly, the transceiver at Alpha Centauri will be executing its own dance due to gravitational forces on it by those three stars engaged in three-body motion over distances of tens of astronomical units. That transceiver motion requires yet more tracking by the Sun's SGL probe, forcing further positional adjustments of millions of kilometers.

*Thus, the use of gravitational lenses to amplify interstellar communication requires that both the transmitter and the receiver move millions of kilometers in synchronized, curlicue motions in space, with an accuracy of roughly a meter, to maintain the alignment with the Sun.* Meeting this requirement is technically possible, but brings obvious failure modes, such as when the spacecraft lose tracking due to a collision with a small Kuiper Belt Object or Oort Cloud comet.

We estimated the propulsive energy requirements for such stationkeeping by building a simple model that included the motion of the Sun due to Jupiter alone (the dominant term). An SGL probe at 550 AU must create its own centripetal acceleration to execute the well-defined "mini-orbit" to track the reflex motion of the Sun. We adopt a benchmark probe mass of 10 kG and lifetime of 1000 yr. The required centripetal force implies a required momentum rate of the exhaust, placing a limit on the product of mass rate and velocity of the exhaust. The benchmark mass (10 kg) and lifetime (1000 years) of the probe constrain the exhaust mass rate at no more than 10 kg per 1000 yr, thus setting a minimum exhaust velocity to achieve the needed propulsive momentum deposition rate.

We find that for a benchmark SGL probe of ~10 kG at 550 AU lasting 1000 years, the stationkeeping force required is ~$10^{-6}$ N and the rate of expenditure of propulsion energy is ~0.1 Watts. For a probe lifetime of a million years, the rate of exhaust mass must be 1/1000 as much, requiring higher exhaust gas speeds to deliver the required momentum, which raises the propulsion energy rate to 50 Watts. Thus, the energy requirements of stationkeeping are less than, or comparable to, the detectable power of transmitted electromagnetic signals. The onboard power budget is not dominated by stationkeeping.

An excellent detailed analysis of the stationkeeping effects, including some not mentioned above, is given by Kerby & Wright (2021). Their independent assessments of stationkeeping are in agreement with those given here, and their work provides more details and effects. To reduce stationkeeping challenges, stars having fewer and less massive companions, both stellar and planetary, are preferred as gravitational lens relay stations.

## 9 DISCUSSION AND SUMMARY

### 9.1 Unique Nature of this Optical SETI Technique

We attempted to detect narrowband ("monochromatic") optical light, pulsed or continuous, coming from a region of the sky where the Sun can serve as a gravitational lens to boost communication sent to Proxima Centauri and Alpha Centauri AB, as proposed by Von Eshleman 1979, Maccone 2011, 2014, 2021, Gillon 2014, Hippke 2020, 2021ab, and Gertz 2021. These regions, at least 550 au from the Sun, are logical locations for Galactic probes and transceivers that are communicating with cohorts in those nearest star systems. The goal was to search for optical, monochromatic light, either sub-second pulses or continuous emission, from any such transmitters. More broadly, we are testing the hypothesis that the Milky Way Galaxy contains robotic probes located at the gravitational lens foci of adjacent stars, motivating a search for laser signals coming from them, intentional or otherwise.

We addressed this science goal by building and operating two objective prism telescopes, operating simultaneously, each equipped with an sCMOS array camera to provide exposure times and cadence of 0.25 s. This dual objective prism system provided the desired spectroscopic and temporal resolution, along with a wide field of view, allowing us to survey the SGL regions for monochromatic light pulses.

There are two useful attributes of this optical search for transients. The spectral resolution of ~4 nm allows monochromatic emission, including multiple emission lines, to stand out prominently and unambiguously compared to broadband imaging of the sky. Also, the time resolution of 0.25 sec is less than the exposure times of 30s to 5 min of typical wide-field imaging surveys for transients. This 0.25 s temporal resolution (and frame rate of 4 fps) allows sub-second pulses of light to stand out prominently against both the light from any host star and the sky background. *A monochromatic pulse lasting a few nanoseconds, microseconds, or milliseconds is detectable in one exposure here, with confirmation from a second telescope, as long as the fluence is above the 200 photon threshold.*

A major consideration in this optical SETI program was to minimize the false positives to avoid their time-consuming follow-up effort. We specifically engineered sufficient spectroscopic resolution to identify and reject astrophysical line-emitting sources and optical pulses composed of a broad spectral energy distribution, such as satellite glints, Cherenkov radiation, and short-lived astrophysical sources such as stellar flares, GRBs, FRBs, and mergers of neutron stars and black holes. The objective prisms provided adequate spectral resolution to retain monochromatic candidates having spectral widths less than the spectral resolution of ~25 Å. Of course, technological communication could be carried by pulses of light that consist of many optical wavelengths. However, we searched for monochromatic light for three reasons: 1. Single-frequency carriers efficiently pack information (as in radio or fiber optic communication), 2. Lasers are proven technology that are nearly monochromatic, and 3. The rejection of false positives that are mostly broadband (e.g. satellite glints of sunlight and astrophysical transients).

We also ruled out several classes of false positives by employing two telescopes, notably elementary particles that directly hit the light sensor. The second telescope was smaller but was sufficient to produce 3-sigma confirmations of 10-sigma candidates found with the larger RASA telescope.

## 9.2 Detection Thresholds in Photon Flux and Source Wattage

We now compute the detection thresholds for sub-second monochromatic pulses of light. During a 0.25 sec exposure, the RASA objective prism system detects monochromatic pulses containing 200 photons within the 2D PSF, with 50% probability. An additional 100 photons raise this to over 90% detection probability. We translate the nominal 200 photon detection threshold to a fluence per unit area using the effective collecting area (including efficiency) of the 0.278-m RASA telescope system, $A_{eff}$ = 0.02 m$^2$. Thus, the detection threshold of 200 photons translates to a fluence threshold in a single exposure of 10000 photons m$^{-2}$ for monochromatic pulses of duration less than 0.25 s. At wavelengths below 400 nm and above 800 nm the quantum efficiency is only ~50% of peak QE (at ~600 nm), thus requiring twice as many photons m$^{-2}$ for detection. Atmospheric extinction of several percent, depending on wavelength, raises this threshold fluence at the top of the Earth's atmosphere by a few percent, not significant here.

*In summary, monochromatic pulses of optical light delivering 10000 photons m$^{-2}$ within 0.25 s would be detected.* Pulses lasting only a nanosecond or up to nearly 1 second duration would be detected. Our search included the 2 deg x 2deg region centered on the SGL regions of Proxima and Alpha Centauri. We observed the SGL region of Proxima Centauri with 88000 exposures lasting a total of 6.1 hr, and the SGL region of Alpha Centauri with 47000 exposures lasting 3.3 hr. By spreading observations over 5 months, parallax would displace the signal location by over 300 arcsec, only rarely overlapping background stars so that the signal would compete only with dark sky. None of the exposures contained a monochromatic pulse above the thresholds described above.

*A diffraction-limited, 1-meter diameter laser with a power of 85 W emitting a wavelength of 500 nm and located 550 au from the Earth would produce a fluence of 10000 photons m$^{-2}$ within 0.25 s, detectable here.* Such a laser would have a beam opening angle of 0.10 arcsec, requiring that it be pointed precisely toward Earth. The beam footprint would be 41000 km in diameter, a few times larger than the Earth. Such a laser beam could hit the entire exposed hemisphere Earth but also could be purposely oriented to avoid hitting the Earth.

Our flux threshold for long-lived monochromatic emission was 3000 photons m$^{-2}$ per sec at Earth from the SGL focus regions of Alpha and Proxima Centauri. We found no such long-lived, monochromatic emission, including astrophysical. We translate that flux threshold of 3000 photons/m$^2$ per sec at the Earth to a power requirement for a hypothetical benchmark laser located at the SGL focus region. *A diffraction-limited, 1-meter diameter laser with a power of only 7 W, located 550 au from Earth and emitting a wavelength of 500 nm, would produce a threshold flux of 3000 photons m$^{-2}$ per sec, detectable here as long-lived monochromatic emission. None was found.*

This required wattage for long-lived emission (7 W) is lower than for the pulses (85 W) because a continuous stream of monochromatic light permits monochromatic flux to accumulate relative to the fluctuations of the sky background that accumulate only as the square root of time. Still, this nominal 7 W diffraction-limited laser would have the same beam opening angle of 0.10 arcsec (as for the nominal pulsed laser), requiring that it be pointed with 0.1 arcsec accuracy toward Earth to be detected. The beam footprint would still be 41000 km in diameter, only a few times larger than the Earth.

Of course, lasers having apertures smaller than 1 meter can be detected. However, their larger diffraction-limited opening beam angle and larger footprint at Earth would require more laser wattage, increasing as the inverse square of laser aperture. The benchmark 7 and 85 W lasers require onboard power such as a radioisotope thermoelectric generator (RTG), a tiny nuclear reactor, or a solar panel ~50 meters across.

### 9.3 Lasers at the SGL Region, Alpha Centauri, and the Milky Way

The non-detection of pulsed and long-lived monochromatic optical light from the SGL regions of Proxima and Alpha Centauri offers yet another domain of SETI parameter space sampled and found empty here. Similarly, no laser emission was found coming directly from Proxima Centauri (Marcy 2021), nor has any narrowband radio emission been found despite intensive and careful observations (Sheikh et al. 2021).

Of course, any probes at the SGL locations could purposely avoid transmitting toward Earth to prevent detection by our telescopes, including spaceborn. Indeed, the Earth never resides exactly on the lines connecting Alpha Cen or Proxima Cen with the Sun, diminishing our hope of fortuitous eavesdropping. From the vantage point of the probe, the Earth would be typically ~5 arcmin away from the Sun. However, the communication beam may be 5 arcmin wide, or the probe may send optical beams toward Earth for reasons we cannot know.

The Sun can be employed for its gravitational amplification at other wavelengths. Communication between probes here and at Proxima or Alpha Cen can be accomplished with lasers operating at infrared, optical, UV, x-ray or gamma-ray frequencies, offering private communication, high bandwidth, and stealth. Such probes communicating with other nearby stars merit our searching (Hippke 2020, 2021ab, Gertz 2021). Galactic communication at UV or x-ray wavelengths would incur less "noise" from the host stars, and would offer greater beam "gain" and bandwidth, favoring searches in those electromagnetic domains.

One may retrospectively evaluate the likelihood of success of this program. Other surveys might have found persistent laser emission if it existed. During the past 80 years, astronomers have carried out many surveys of the entire sky for objects that emit emission lines such as planetary nebulae, HII regions, T Tauri stars, Be stars, M dwarf flare stars, active galactic nuclei of various types (including quasars), and high redshift Lyman alpha emitters, to name a few. The entire sky has been searched, including along the plane and out of the plane of the Milky Way. Most of these searches proceeded by employing wide-field, low-resolution optical spectroscopy or narrow-band photometry, e.g., with objective prism telescopes or with direct imaging using narrow-band filters, permitting the identifications of thousands of line emitting objects documented in catalogs. Follow-up spectroscopy permits characterization and modeling. Objects with persistent line emission having broadband brightness as faint as $17^{th}$ magnitude were routinely detected and studied. But no technological signals emerged from past surveys.

Remarkably, wide-field optical surveys within and outside the plane of the Milky Way revealed objects exhibiting only one dominant emission feature, notably H-alpha or [OIII], at low Doppler shifts less than 100 km s$^{-1}$. If an emission line did not correspond to H-alpha or [OIII], the wavelength oddity might have motivated follow-up spectroscopy and high spatial resolution imaging. The astronomers doing the survey may fail to notice unexpected phenomena or ignore them due to time constraints. The multi-dimensional parameter space of SETI is vast (Wright et al. 2018). Nevertheless, some astronomers are vigilant for the unexpected and follow-up with further investigation. Indeed, high redshift galaxies that emit Lyman alpha exhibit emission at arbitrary wavelengths in the observatory frame of reference, attracting numerous follow-up observational studies to elucidate the nature of the object. Laser emission that exhibits no continuum or other emission lines would stand out in Lyman alpha surveys at optical wavelengths. These sky surveys found no persistent laser emission, *constituting a vast, implicit SETI non-detection.* The emission, even if persistent, would cause a short signal, recurring rarely if at all, as seen from the Earth. The duration of the signal may last no more than a few seconds, requiring the survey observations to occur at the right time and at the right wavelength.

In addition, we have obtained high resolution spectra of the nearest 3000 stars and searched them for laser emission lines (Reines & Marcy 2002, Tellis & Marcy 2017). We found none. Thus the *a priori* probability that this present survey would identify such persistent emission was probably low from the start (see Section 5 of Marcy 2021), but difficult to quantify.

Similarly, SETI surveys at other wavelength domains (e.g., radio, IR) suffer from implicit competition from prior all-sky surveys at those wavelengths. Even broadband surveys can detect narrowband signals above threshold flux levels, warranting follow-up spectroscopy. We must acknowledge the significance of the growing collection of non-detections in SETI, pursued either intentionally or implicitly, in past astrophysics surveys. The probability is increasing that either technological transmission beams fill a small fraction of the volume of space or that the transmitters are purposely pointed away from Earth. Or both.

We plan to search for monochromatic pulses from the SGL regions of other nearby stars, notably tau Ceti, epsilon Eri, delta Pav, Barnard's star, GJ 406, and Sirius. We are also completing a search for monochromatic pulses from the center and disk of the Milky Way Galaxy and from other galaxies using superior cameras, namely a QHY600M.

## 10   ACKNOWLEDGMENTS

This paper was motivated by discussions with and papers by Michael Hippke (see Hippke 2020, 2021a, 2021b,2021c). The work benefitted from valuable communications with Beatriz Villarroel, Susan Kegley, John Gertz, Brian Hill, Dan Werthimer, Ben Zuckerman, Ariana Paul, Franklin Antonio, Claudio Grimaldi, Carmella Martinez, Roger Bland, Saundra Norton, David Rowe, Michael Garrett, Rudolf Baer, Lars Mattsson, Alok Gupta, David Brin. We thank the team at Space Laser Awareness for outstanding technical help. We thank the team at Finger Lakes Instrumentation who made the KL400 camera, the ZWO team who made the ASI1600 camera, and TheSkyX team for software that operates the telescope and ZWO camera.

DATA AVAILABILITY

This paper is based on raw CMOS sensor images obtained with Space Laser Awareness double objective prism telescopes. The 135000 images are 8.4 Mb each, located on a peripheral disk with 1.1 terabytes of storage, not online. All images are available to the public upon the request of G.M., and a transfer method must be identified.

This paper has been typeset from Microsoft WORD document prepared by the author.